 \documentclass[final,5p,times,twocolumn,authoryear]{elsarticle}

\usepackage{amsmath, amssymb}
\usepackage{lipsum}
\usepackage{array}
\usepackage{graphicx, hyperref}
\usepackage{booktabs}
\usepackage{pgfplots}
\usepackage{helvet}
\usepackage{ifthen}
\usepackage{supertabular}
\usepackage{longtable}
\journal{Astronomy $\&$ Computing}

\begin{document}

\begin{frontmatter}

%% Title, authors and addresses

%% use the tnoteref command within \title for footnotes;
%% use the tnotetext command for theassociated footnote;
%% use the fnref command within \author or \affiliation for footnotes;
%% use the fntext command for theassociated footnote;
%% use the corref command within \author for corresponding author footnotes;
%% use the cortext command for theassociated footnote;
%% use the ead command for the email address,
%% and the form \ead[url] for the home page:
%% \title{Title\tnoteref{label1}}
%% \tnotetext[label1]{}
%% \author{Name\corref{cor1}\fnref{label2}}
%% \ead{email address}
%% \ead[url]{home page}
%% \fntext[label2]{}
%% \cortext[cor1]{}
%% \affiliation{organization={},
%%            addressline={}, 
%%            city={},
%%            postcode={}, 
%%            state={},
%%            country={}}
%% \fntext[label3]{}

\title{Galaxy Morphological Classification with Manifold Learning}

\author[first]{Vasyl Semenov}
\author[first,second]{Vitalii Tymchyshyn}
\author[first]{Volodymyr Bezguba}
\author[third,fourth]{Maksym Tsizh}
\author[first]{Andrii Khlevniuk}

\affiliation[first]{organization={Kyiv Academic University},
            addressline={Vernadsky blvd. 36}, 
            city={Kyiv},
            postcode={UA-03142}, 
            %state={},
            country={Ukraine}}
\affiliation[second]{organization={Bogolyubov Institute for Theoretical Physics},
            addressline={Metrolohichna 14-b}, 
            city={Kyiv},
            postcode={UA-02000}, 
            %state={},
            country={Ukraine}}
\affiliation[third]{organization={Dipartimento di Fisica e Astronomia, Universitá di Bologna},
            addressline={Via Gobetti 92/3}, 
            city={Bologna},
            postcode={40121}, 
            %state={},
            country={Italy}}
\affiliation[fourth]{organization={Astronomical Observatory of Ivan Franko National University of Lviv},
            addressline={Kyryla i Methodia str. 8}, 
            city={Lviv},
            postcode={79005}, 
            %state={},
            country={Ukraine}}

\begin{abstract}
We address the problem of morphological classification of galaxies from the Galaxy Zoo DECaLS dataset using classical machine learning techniques. Our approach employs a dimensionality reduction method followed by a classical classifier to categorize galaxies based on shape (cigar/in-between/ round; edge-on/face-on) and texture (smooth/featured).
We evaluate various dimensionality reduction methods, including Locally Linear Embedding (LLE), Isomap, Uniform Manifold Approximation and Projection (UMAP), t-SNE, and Principal Component Analysis (PCA). Our results demonstrate that most classical classifiers achieve their highest performance when combined with LLE, attaining accuracy comparable to that of simple neural networks. Moreover, in the case of shape classification, the three-dimensional representation remains interpretable, in contrast to the commonly observed loss of interpretability following nonlinear transformations.
Additionally, we explore dimensionality reduction followed by k-means clustering to assess whether the data exhibits a natural tendency toward a specific number of clusters. We evaluate clustering performance using silhouette, elbow, Dunn, and Davies-Bouldin scores. While the Davies-Bouldin score indicates a slight preference for four clusters—closely aligning with classifications made by human astronomers—the other metrics do not support a distinct clustering structure.
\end{abstract}

\end{frontmatter}

%\tableofcontents

%% \linenumbers

%% main text

\section{Introduction}
\label{intro}
The morphological galaxy classification is one of the oldest classification problems in astrophysics. Since the discovery of existing galaxies as separate structures of matter, people have noticed that they can be divided into several distinct morphological classes. The modern era of deep sky surveys with high-resolution imagery has brought variegation to this problem. Consequently, a significant number of attempts have emerged to apply different machine learning (ML) methods to help resolve this problem.

The Galaxy Zoo (GZ) project has greatly facilitated exploring the possibilities of machine learning for the classification of galaxies. The Galaxy Zoo community did a fantastic job forming a large, clean, and well-maintained human-labeled database of galaxy images from several surveys. The databases of the project have quickly become a test polygon of various ML models, from the simplest ones such as k-nearest neighbors \citep{Shamir09}, support vector machines (with \citep{Barchi2016} and without feature extraction \citep{Polsterer12}), and simple versions of neural networks \citep{Banerji10}, to much more complicated ones, such as convolutional neural networks (\citep{Dieleman2015}, \citep{DomnguezSnchez2018}, \citep{Vavilova22}, in particular, efficientNet in \citep{Kalvankar20}), and autoencoders \citep{Xu23}. With time, the variety of methods applied to the morphological classification of GZ galaxies grew rapidly: the vision transformer was used to this end in \cite{Lin21}, and adversarial deep learning in \cite{iprijanovi2022}. A comparison of deep learning techniques on GZ data was performed in \citep{Fielding2021} and with the release of Zoobot by Walmsley and colleagues \citep{Walmsley2023} one can fine-tune its own deep network to classify the GZ galaxies. The linear discriminant analysis technique was used by F. Ferrari et al. \citep{Ferrari2015} to excrete morphological parameters and classify galaxies of early GZ (based on SDSS observation) output with an accuracy of ~90\%. For the same purpose, \citep{Rosa2018} exploits the gradient pattern analysis for the larger data set with slightly better results. Comparison of the classification of the GZ dataset using a non-parametric approach and deep learning was performed in \citep{Barchi2020}. The paper \citep{Kolesnikov2023} can be recognized as an interesting "bridge" between supervised and unsupervised machine learning methods, as it utilizes a hybrid approach for the classification of GZ images. In addition, a blend of computer and human classification was used for GZ classification task \citep{Beck2018}.

The important question of the robustness of the results of deep learning algorithms classification is addressed in \citep{iprijanovi2021} and the bias in galaxy morphology classification is revealed in \citep{CabreraVives2018}.

In this work, we use one of the latest releases of the Galaxy Zoo project - Galaxy Zoo DECaLS (GZ DECaLS) \citep{bib:GZDECaLS}. It was built on the Dark Energy Camera Legacy Survey data and contains the classification of more than 300,000 galaxies. This particular database was also subjected to numerous endeavors of machine-learning classification of galaxies. Consider some of the unsupervised models. The contrastive learning was used in \citep{Wei22}. In \citep{Fielding22}, the authors used a convolutional autoencoder feature extractor before applying k-means, fuzzy c-means, and agglomerative clustering to the extracted data features to cluster them. In a similar way, a blend of vector-quantized variational autoencoder with hierarchical clustering was applied in \citep{Cheng2021} to the SDSS galaxies unsupervised classification. Lowering dimensionality was performed with this dataset in \citep{Waghumbare24} which was followed by the automated clustering algorithm. One of the methods of family manifold learning, t-Distributed Stochastic Neighbor Embedding (t-SNE), was applied \citep{Dai18} for visualization of the last layers of the CNN network, that classified GZ2 data.

On the other hand, the manifold learning methods of dimensionality reduction, particularly the Locally Linear Embedding, have not yet been used in any morphological classification in astrophysics. Instead, it was utilized when solving problems of galaxy \citep{Vanderplas09}, quasar \citep{Jankov20}, stellar \citep{Daniel11}, \citep{Yude13}, and protostellar \citep{Ward16} spectra classification, as well as binary light curves classification \citep{Bodi21}. Thus, broadening the scope of its application as a denoising and feature extracting tool appears promising. 

This paper is organized as follows. In Section \ref{methods}, we discuss our workflow, including data preparation, the methods employed, the pipelines, and the libraries used. We also explain the architecture of the neural networks utilized for performance comparison. In Section \ref{results}, we present the results of the classification and clustering and argue for the superior performance of the LLE method. Section \ref{interp} focuses on interpreting and visualizing the reduced data in the case of three dimensions. Finally, we provide a brief conclusion in Section \ref{conc}.

The code to reproduce all our results can be found in a public repository~\cite{our_repo}.

\section{Methods}
\label{methods}

\subsection{Data Preparation}

In the current contribution, we used the GZD-5 database~\cite{bib:GZDECaLS} which contains $424 \times 424$ images of galaxies along with classification provided by volunteers.
First, the data were curated according to classification schemes (see table~\ref{tbl_data_curation}).
Please note, that a single image can belong to several classes if they come from different classification problems, thus the total number of images is just $55137$.

\begin{table}[ht]
\centering
\begin{tabular}{|l|rr|r|}
\hline
class & class fraction & voters & images\\
\hline
round & $> 0.8$ & $> 10$ & $8975$\\
in-between & $> 0.8$ & $> 10$ & $13725$\\
cigar & $> 0.8$ & $> 10$ & $4491$\\
\hline
edge-on & $> 0.8$ & $> 10$ & $5120$\\
face-on & $> 0.8$ & $> 10$ & $24493$\\
\hline
smooth & $> 0.8$ & $> 10$ & $13558$\\
featured & $< 0.2$ & $> 10$ & $12753$\\
\hline
unsupervised & N/A & N/A & $55137$\\
\hline
\end{tabular}
\caption{Data filtering conditions.}
\label{tbl_data_curation}
\end{table}

\begin{figure}[hbt]
\centering
\setlength{\tabcolsep}{2pt}
\begin{tabular}{rl}
\includegraphics[width=0.45\linewidth]{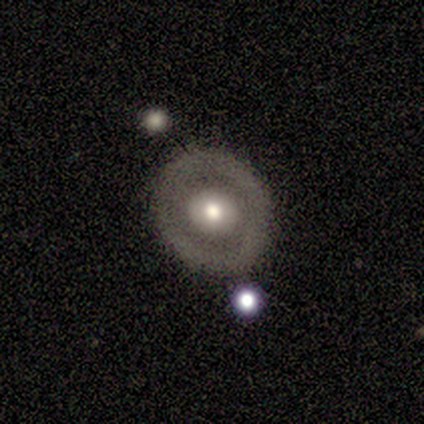}&
\includegraphics[width=0.45\linewidth]{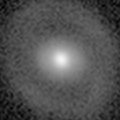}\\
\includegraphics[width=0.45\linewidth]{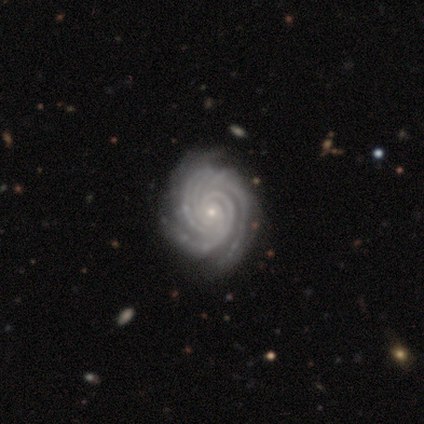}&
\includegraphics[width=0.45\linewidth]{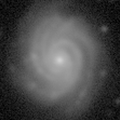}\\
\includegraphics[width=0.45\linewidth]{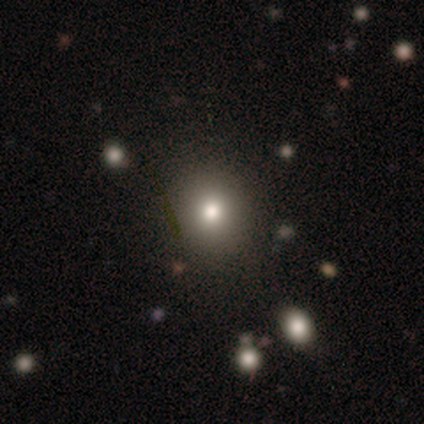}&
\includegraphics[width=0.45\linewidth]{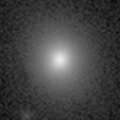}\\
\end{tabular}
    \caption{Original photos from GZD-5 database (left) and their cropped and grayscaled versions (right)}
    \label{fig:galaxies_example}
\end{figure}

All images of the curated dataset were preprocessed for efficiency: we extracted the $120 \times 120$ central part and converted it to grayscale.
An example of this pre-processing is shown in~\ref{fig:galaxies_example}.

\begin{figure}[ht]
\centering
\setlength{\tabcolsep}{3pt}
\begin{tabular}{lr}
\includegraphics[width=0.45\linewidth]{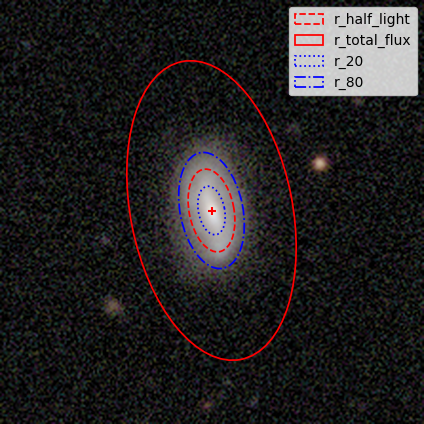} &
\includegraphics[width=0.45\linewidth]{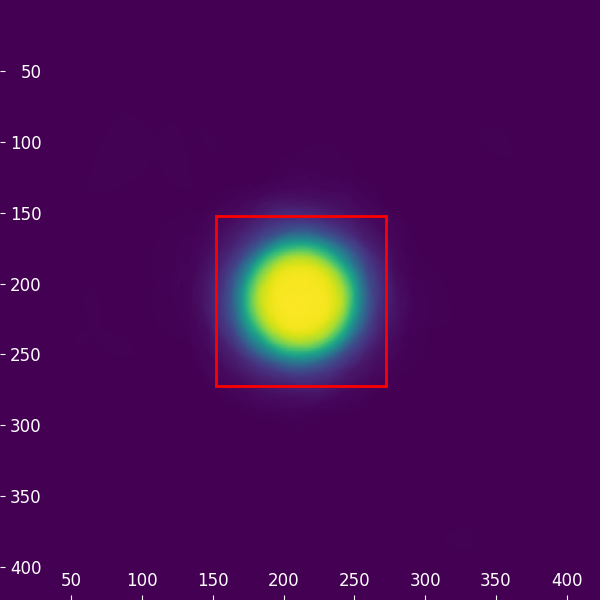}
\end{tabular}
    \caption{Example of the {\it petrofit}-processed image (left) and an overlay of  $r_{80}$ ellipses inside $120\times 120$ square (right).}
    \label{fig:petrofit}
\end{figure}

The crop size was chosen based on the Petrosian $r_{80}$ radii of the images.
We employed the \textit{petrofit} library \cite{Geda2022} to calculate the appropriate radii on a random subsample ($700$ images) of the curated dataset.
The $95\%$ confidence interval for the mean value of the $r_{80}$ parameter was estimated as $53 \pm 2$ pixels that corresponds to window size $106 \pm 4$ pixels.
To round up we took the window size to be $120 \times 120$.
An example of the processing results is shown in~\ref{fig:petrofit}.

Moreover, we used $r_{80}$ of the galaxies to create an overlay image that shows the probability of a certain pixel to fall into the $r_{80}$ range.
The result is presented in the same figure~\ref{fig:petrofit}.

\subsection{Pipeline}

The following pipeline was used: a dimensionality reduction method followed by a classical ML classification.
We were mostly interested in finding the best dimensionality reduction method that was measured as the accuracy of the subsequent classification for the most complex classification scheme of $3$ classes.

In this work, we compared the LLE, Isomap, t-SNE, PCA, and UMAP algorithms.
We used the \textit{scikit-learn} library to calculate LLE and Isomap, while t-SNE, PCA, and UMAP were calculated by the \textit{cuml} library (see \textit{rapids.ai}).
The latter performs computation on a GPU that significantly speeds up the processing, but its implementation of the t-SNE employs the Barnes-Hut algorithm and is restricted to $2$ dimensions only.
The classification methods we considered include logistic regression, support vector classifier, and decision tree implemented in the \textit{scikit-learn} library.

As input to the pipeline, we used the preprocessed images described above. Each image is treated as a data point in a multi-dimensional space, where each pixel serves as a separate coordinate.

\subsection{Parameter Optimization}

For each of the considered dimensionality reduction and classification methods, we performed Bayesian optimization using the {\it optuna} library \cite{optuna}.
We conducted $50$ trials for each combination, with parameters constrained within the appropriate ranges, along with various classification methods.

\begin{table}[ht]
\centering
\begin{tabular}{lrrr}
\toprule
method & components & neighbors & perplexity \\
\midrule
LLE & 2--200 & 2--50 & N/A \\
UMAP & 2--200 & 5--50 & N/A \\
Isomap & 2--200 & 10--50 & N/A \\
PCA & 2--200 & N/A & N/A \\
t-SNE & 2 & 18--153 & 5--50 \\
\bottomrule
\end{tabular}
\caption{Parameters ranges used for optimization by {\it optuna}. ``N/A''\,--- not applicable.}
\label{tbl_params}
\end{table}

Please note that some additional restrictions may apply in the code. For example, the number of components for t-SNE should be greater than three times the perplexity.

As a measure of performance, we considered the accuracy of the subsequent classification obtained through $5$-fold cross-validation on a sample of $5,000$ images from the curated and pre-processed dataset. We performed such a search for each case of classification separately and later used the most optimal parameters for each case correspondingly.  See table~\ref{tbl_params} for results for round/in-between/
cigar classification. In each case, the LLE method appeared to be the most efficient.

\subsection{Comparison with Neural Networks}

\begin{figure}
    \centering
    \includegraphics[width=0.5\linewidth]{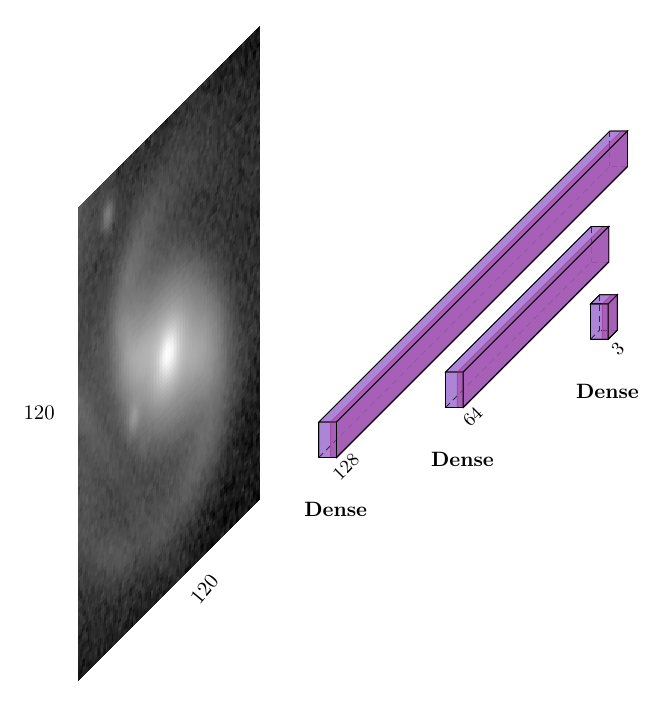}
    \caption{Fully-connected neural network architecture. Total number of parameters $16\, 771$, all tunable.}
    \label{fig:fcnn}
\end{figure}

To put our approach into the perspective of modern NN-based classification techniques, we created two simple neural networks.
The first one is a simple three-layer perceptron (fully-connected neural network, FCNN) with the architecture shown in the figure~\ref{fig:fcnn}.
The first two layers have \textit{ReLU} as an activation function,  while the last one uses \textit{softmax}.

\begin{figure*}[t]
    \centering
    \includegraphics[width=\linewidth]{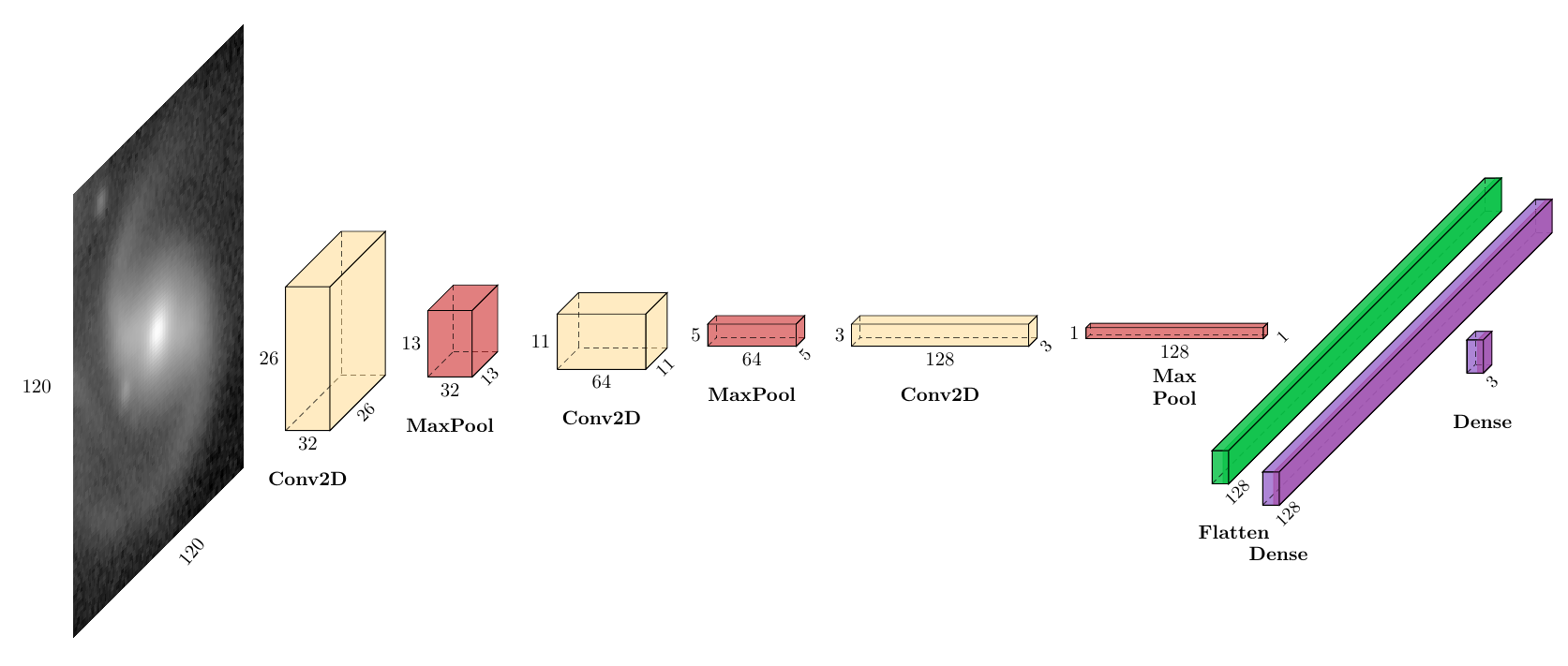}
    \caption{CNN architecture. Total number of parameters $109\,571$, all tunable.}
    \label{fig:cnn_architecture}
\end{figure*}

The second one is a simple convolutional neural network with an architecture shown in the figure~\ref{fig:cnn_architecture}.
All convolutional layers employ \textit{ReLU} as their activation function.
As for the last two fully-connected layers, the first one employs \textit{ReLU}, while the last one uses \textit{softmax}. 

For both neural networks, we used a sample of $20,000$ images, batch size $32$.
The train/test split was performed as $70/30$, with $14,000$ images for training and $6,000$ for testing.
We used \textit{adam} optimizer and \textit{categorical crossentropy} as the loss function.
For additional control, we later re-evaluated the pre-trained network on a different random sample of $20,000$ images.

\section{Results}
\label{results}

\subsection{Parameters Estimation}

Our tests suggest that the best-performing dimensionality reduction method is independent of the subsequent classification and the best method is found to be LLE. But it should be noted that there is only a small margin between the first and second places. The table~\ref{tbl_lle_params} presents the best-fine-tuned parameters for logistic regression.

\begin{table}[ht]
\begin{tabular}{rlrrr}
\toprule
accuracy & method & components & neighbors & perplexity \\
\midrule
0.93 & LLE & 138 & 10 & N/A \\
0.92 & UMAP & 82 & 26 & N/A \\
0.88 & Isomap & 87 & 10 & N/A \\
0.61 & PCA & 63 & N/A & N/A \\
0.54 & t-SNE & 2 & 102 & 33 \\
\bottomrule
\end{tabular}
\caption{Fine-tuned parameters for each of the dimensionality reduction methods used. The classification method in the pipeline is logistic regression, the classification case is round/in-between/cigar}
\label{tbl_lle_params}
\end{table}

Additionally, the rest of the leaderboard depends on the classification method used.
If we were to use the support vector classifier with the RBF kernel, the second place would have been taken by the PCA, despite its relatively poor performance with the logistic regression classifier. We speculate this is due to the fact that the highly non-linear kernel of the SVC compensates for the excessively restricted PCA.

The t-SNE method demonstrates relatively poor performance due to the restriction to only $2$ coordinates.
Since we used the GPU-based implementation from \textit{rapids.ai}, which implements the Barnes-Hut algorithm and allows for only 2 dimensions. We expect the results from \textit{scikit-learn} implementation to be much better, but processing time is estimated to be a few days, which makes this method unfeasible.

\subsection{Classification Performance}

After parameter optimization was performed, we tested our pipeline on a subsample of $20,000$ images for three different classification schemes.
For dimensionality reduction, we used LLE as the best performer, while classification was done by logistic regression.
The train/test splitting was $70/30$, with $14,000$ images in the training set and $6,000$ in the test set.

\begin{figure}[htb]
\centering    
\setlength{\tabcolsep}{15pt}
\begin{tabular}{ccc}
\includegraphics[width=0.2\linewidth]{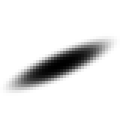}&
\includegraphics[width=0.2\linewidth]{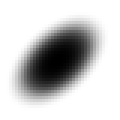}&
\includegraphics[width=0.2\linewidth]{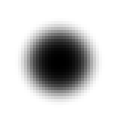}
\end{tabular}
    \caption{Round/in-between/cigar galaxy types from (\cite{bib:GZDECaLS}).}
    \label{fig:how_rounded}
\end{figure}

The first type of classification is round/in-between/cigar scheme presented in figure~\ref{fig:how_rounded}.
One could say that the galaxies are classified based on their apparent eccentricity.

The results can be presented in the form of confusion matrix~\ref{fig:conf_matr_rounded} and table~\ref{tab:class1report}.
It should be noted that the classification performance weakly depends on the parameters.
Varying the number of neighbors and dimensionality over a wide range results in only a $1-2\%$  difference between the obtained accuracy and the best-reported result.

\begin{figure}[htb]
\centering
\begin{tikzpicture}
\begin{axis}[
    scale=0.7,
    colormap={whitenavy}{rgb=(1,1,1) rgb=(0.03,0.19,0.42)},
    enlargelimits=false,
    axis on top,
    xtick={0,1,2}, xticklabels={round, in-between, cigar},
    ytick={0,1,2}, yticklabels={\vphantom{g}round, \vphantom{g}in-between, cigar},
    y tick label style={rotate=90},
    xlabel={Predicted},
    ylabel={True},
    ylabel style={yshift=-3pt},
    xlabel style={yshift=-10pt},
    xmin=-0.5, xmax=2.5,
    ymin=-0.5, ymax=2.5,
    axis equal image,
    tick style={draw=none},
    visualization depends on=\thisrow{C} \as \x,
    every node near coord/.append style={
        anchor=center,
        font=\large
    }
]
\addplot [matrix plot, mesh/cols=3, nodes near coords, point meta=explicit, 
        visualization depends on=\thisrow{C} \as \x,
        nodes near coords={
        \pgfmathsetmacro\mycolor{\x<1200 ? "black" : "white"}
        \textcolor{\mycolor}{\pgfmathprintnumber{\x}}
        }
    ] table[meta=C] {
x y C
0 0 1875
0 1 96
0 2 4
1 0 109
1 1 2882
1 2 70
2 0 3
2 1 43
2 2 918
};
\end{axis}
\end{tikzpicture}
    \caption{Confusion matrix for round/in-between/cigar scheme. Pipeline: LLE + logistic regression.}
    \label{fig:conf_matr_rounded}
\end{figure} 

\begin{table}[htb]
\centering
\begin{tabular}{lrrrr}
\toprule
 & precision & recall & f1-score & support \\
\midrule
round & 0.95 & 0.94 & 0.95 & 1,987 \\
inbetween & 0.94 & 0.95 & 0.95 & 3,021 \\
cigar & 0.95 & 0.93 & 0.94 & 992 \\
\hline
accuracy & 0.95 & - & - & - \\
macro avg & 0.95 & 0.94 & 0.94 & 6,000 \\
weighted avg & 0.95 & 0.95 & 0.95 & 6,000 \\
\bottomrule
\end{tabular}
\caption{Classification report for round/in-between/cigar scheme. Pipeline: LLE + logistic regression.}
\label{tab:class1report}
\end{table}

Comparing this case to neural networks, CNN performs a bit better with weighted averages: precision $98\%$, recall $98\%$, and f1-score also $98\%$, while the multilayer perceptron (fully-connected neural network, FCNN) exhibits the same performance: precision $95\%$, recall $95\%$, and f1-score also $95\%$.

\begin{figure}[htb]
\centering    
\setlength{\tabcolsep}{25pt}
\begin{tabular}{cc}
\includegraphics[width=0.2\linewidth]{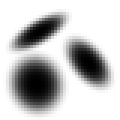}&
\includegraphics[width=0.2\linewidth]{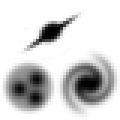}
\end{tabular}
    \caption{Smooth/featured galaxy types from (\cite{bib:GZDECaLS}).}
    \label{fig:featured}
\end{figure}

The second scheme we considered was the smooth/featured scheme, as shown in figure~\ref{fig:featured}.
Using the same parameters as in the previous case, we get results presented in figure~\ref{fig:conf_matr_featured} and table~\ref{tab:class2report}.

\begin{figure}[htb]
\centering
\begin{tikzpicture}
\begin{axis}[
    scale=0.5,
    colormap={whitenavy}{rgb=(1,1,1) rgb=(0.03,0.19,0.42)},
    enlargelimits=false,
    axis on top,
    xtick={0,1}, xticklabels={smooth, featured},
    ytick={0,1}, yticklabels={\vphantom{g}smooth, \vphantom{g}featured},
    y tick label style={rotate=90},
    xlabel={Predicted},
    ylabel={True},
    ylabel style={yshift=-3pt},
    xlabel style={yshift=-10pt},
    xmin=-0.5, xmax=1.5,
    ymin=-0.5, ymax=1.5,
    axis equal image,
    tick style={draw=none},
    visualization depends on=\thisrow{C} \as \x,
    every node near coord/.append style={
        anchor=center,
        font=\large
    }
]
\addplot [matrix plot, mesh/cols=2, nodes near coords, point meta=explicit, 
        visualization depends on=\thisrow{C} \as \x,
        nodes near coords={
        \pgfmathsetmacro\mycolor{\x<1200 ? "black" : "white"}
        \textcolor{\mycolor}{\pgfmathprintnumber{\x}}
        }
    ] table[meta=C] {
x y C
0 0 2428
0 1 653
1 0 652
1 1 2267
};
\end{axis}
\end{tikzpicture}
    \caption{Confusion matrix for smooth/featured classification scheme. Pipeline: LLE + logistic regression.}
    \label{fig:conf_matr_featured}
\end{figure}

\begin{table}[htb]
\centering
\begin{tabular}{lrrrr}
\toprule
 & precision & recall & f1-score & support \\
\midrule
smooth & 0.86 & 0.85 & 0.86 & 3,103 \\
featured & 0.85 & 0.85 & 0.85 & 2,897 \\
\hline
accuracy & 0.85 & - & - & - \\
macro avg & 0.78 & 0.85 & 0.85 & 6,000 \\
weighted avg & 0.85 & 0.85 & 0.85 & 6,000 \\
\bottomrule
\end{tabular}
\caption{Classification report for smooth/featured classification scheme. Pipeline: LLE + logistic regression.}
\label{tab:class2report}
\end{table}

Once again, we observe that the results are only weakly dependent on the parameters. Therefore, performing a separate \textit{optuna}-based optimization in this case will not lead to significant improvement.
Compared to the neural networks, we obtain better results for the CNN: $98\%$ for weighted precision, recall, and f1-score; the perceptron exhibits lower performance with weighted precision $84\%$ and $81\%$ for recall and f1-score. 

\begin{figure}[htb]
\centering   
\setlength{\tabcolsep}{25pt}
\begin{tabular}{cc}
\includegraphics[width=0.2\linewidth]{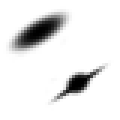}&
\includegraphics[width=0.2\linewidth]{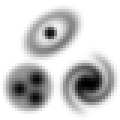}
\end{tabular}
    \caption{Edge-on/face-on galaxy types from (\cite{bib:GZDECaLS}).}
    \label{fig:edge_on_disk}
\end{figure}

The last classification scheme considered is edge-on/face-on (see figure~\ref{fig:edge_on_disk}).
Once again, we use the same parameters as in the previous cases to obtain the results presented in figure~\ref{fig:conf_matr_disk} and table~\ref{tab:class3report}.

\begin{figure}[htb]
\centering
\begin{tikzpicture}
\begin{axis}[
    scale=0.5,
    colormap={whitenavy}{rgb=(1,1,1) rgb=(0.03,0.19,0.42)},
    enlargelimits=false,
    axis on top,
    xtick={0,1}, xticklabels={edge-on, face-on},
    ytick={0,1}, yticklabels={\vphantom{g}edge-on, \vphantom{g}face-on},
    y tick label style={rotate=90},
    xlabel={Predicted},
    ylabel={True},
    ylabel style={yshift=-3pt},
    xlabel style={yshift=-10pt},
    xmin=-0.5, xmax=1.5,
    ymin=-0.5, ymax=1.5,
    axis equal image,
    tick style={draw=none},
    visualization depends on=\thisrow{C} \as \x,
    every node near coord/.append style={
        anchor=center,
        font=\large
    }
]
\addplot [matrix plot, mesh/cols=2, nodes near coords, point meta=explicit, 
        visualization depends on=\thisrow{C} \as \x,
        nodes near coords={
        \pgfmathsetmacro\mycolor{\x<1200 ? "black" : "white"}
        \textcolor{\mycolor}{\pgfmathprintnumber{\x}}
        }
    ] table[meta=C] {
x y C
0 0 880
0 1 103
1 0 156
1 1 4861
};
\end{axis}
\end{tikzpicture}
    \caption{Confusion matrix for edge-on/face-on classification scheme. Pipeline: LLE + logistic regression.}
    \label{fig:conf_matr_disk}
\end{figure} 

\begin{table}[htb]
\centering
\begin{tabular}{lrrrr}
\toprule
 & precision & recall & f1-score & support \\
\midrule
edge-on & 0.90 & 0.88 & 0.89 & 1,036 \\
face-on & 0.98 & 0.98 & 0.98 & 4,964 \\
\hline
accuracy & 0.96 & - & - & - \\
macro avg & 0.94 & 0.93 & 0.94 & 6,000 \\
weighted avg & 0.96 & 0.96 & 0.96 & 6,000 \\
\bottomrule
\end{tabular}
\caption{Classification report for edge-on/face-on classification scheme. Pipeline: LLE + logistic regression.}
\label{tab:class3report}
\end{table}

This case is no exception in terms of stability under parameter variations. The performance exhibits only a weak dependence on dimensionality and the number of neighbors.

Regarding comparison with the neural networks, it is a typical picture with CNN being better ($98\%$ weighted precision, recall, and f1-score), while FCNN scores lower ($95\%$ weighted precision, recall, and f1-score).

\subsection{Clusterization}

We explored dimensionality reduction followed by k-means clustering to assess whether the data exhibits a natural tendency toward a specific number of clusters.

The clustering pipeline differs slightly from the one used for classification. First, we sampled $20,000$ random images from the curated dataset, ignoring their classes (the maximum dataset size that Google Colab can handle). The LLE method was employed for dimensionality reduction, using the same parameters as in the classification tasks. The resulting embedding was then used as the dataset.

\begin{figure}[htb]
\centering    
\includegraphics[width=\linewidth]{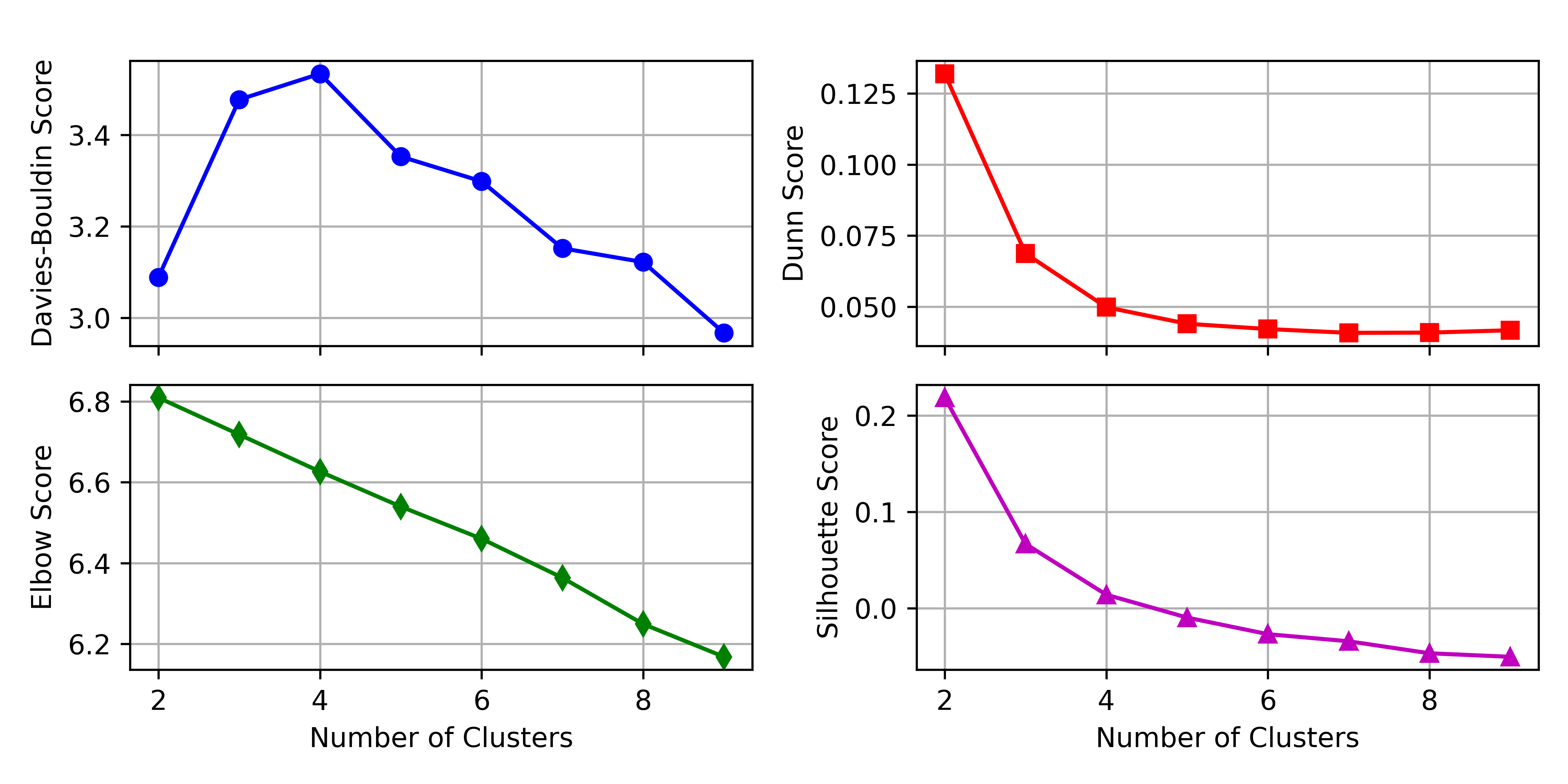}
\caption{Different scores for clustering. Pipeline: LLE + k-means.}
\label{fig_clust_scores}
\end{figure}

The second step involved obtaining a $1,000$ datapoint subsample of the LLE-processed images, performing k-means clustering, and estimating several different scores: silhouette, elbow, Dunn, and Davies-Bouldin. This procedure was repeated $1,000$ times to collect statistics and generate mean curves for each score.
The results are presented in figure~\ref{fig_clust_scores}.

It can be observed that the Davies-Bouldin score indicates a slight preference for four clusters, closely aligning with classifications made by human astronomers (elliptical, spiral, lenticular, and irregular). The other metrics do not support a distinct clustering structure.
We suggest that further research is needed to find a certain pipeline capable of unsupervised galaxy discrimination.

\section{Interpretability}
\label{interp}

A thorough analysis revealed that the performance of the supervised classification exhibits a weak dependence on the dimension of the dimensionality reduction algorithms. Consequently, we leveraged the visualization capabilities of the 3D space to interpret the components of the galaxy images following the dimensionality reduction phase.

\begin{figure}[htb]
\centering
\includegraphics[width=0.75\linewidth]{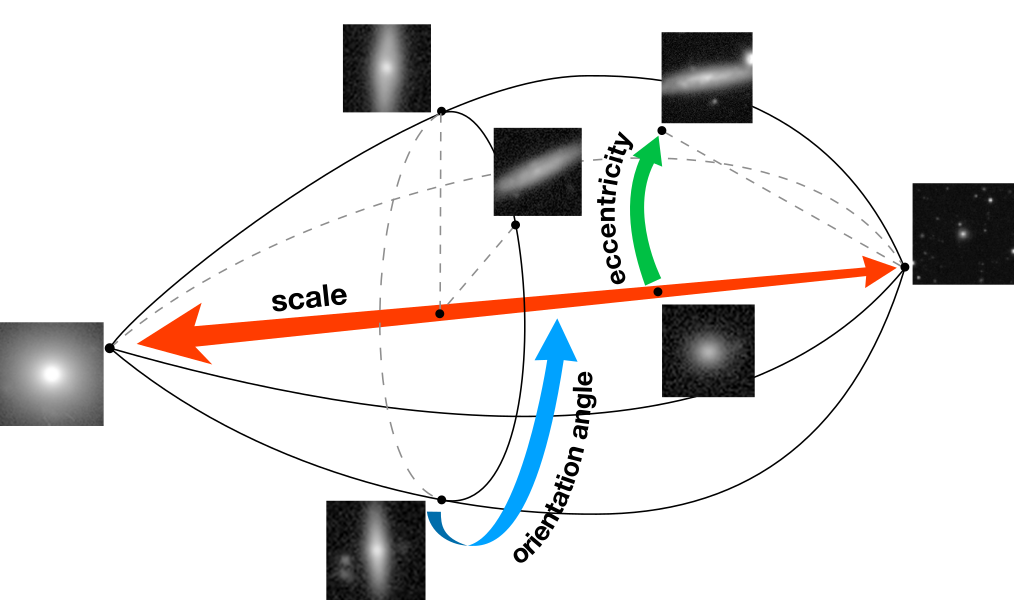}
\caption{Manifold scheme}
\label{fig_manifold_scheme}
\end{figure}

\begin{figure*}[p]
\newlength{\myimageheight}
\setlength{\myimageheight}{0.14\paperheight}
\newcolumntype{C}[1]{>{\centering\arraybackslash}p{#1}}
\centering
%\caption{A long table spanning both columns}
%\vspace{-1.5cm}
\begin{supertabular}{|C{0.45\linewidth}|C{0.45\linewidth}|}
\hline
\includegraphics[height=\myimageheight]{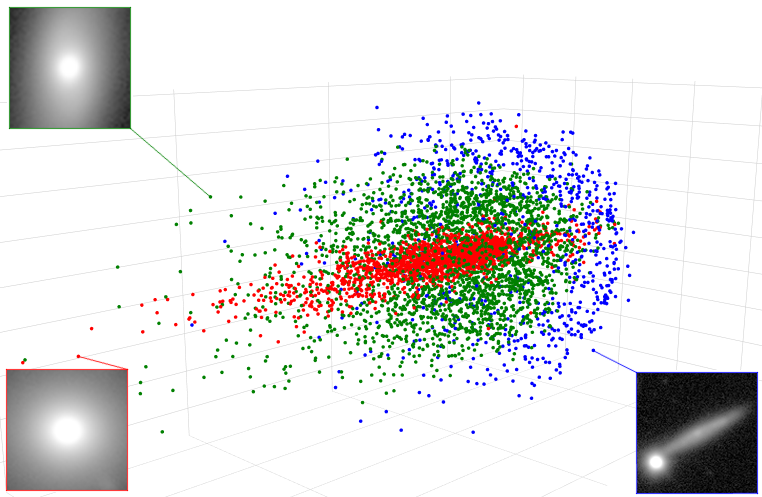} & \includegraphics[height=\myimageheight]{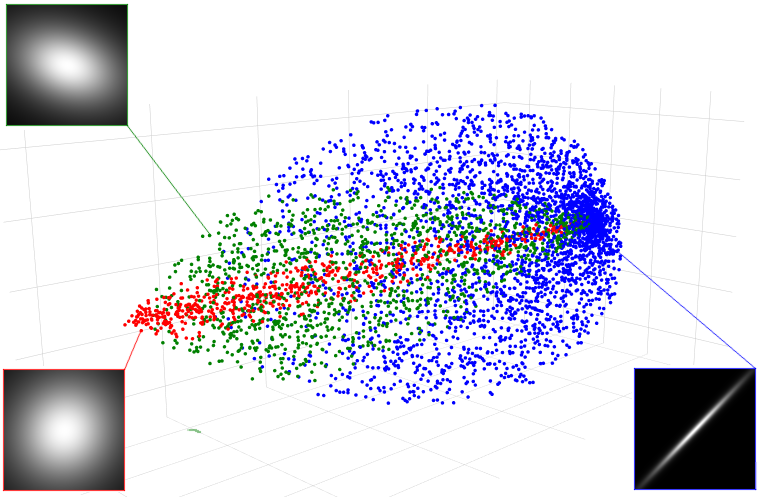} \\
GZD-5 PCA & synthetic dataset PCA \\
\hline
\includegraphics[height=\myimageheight]{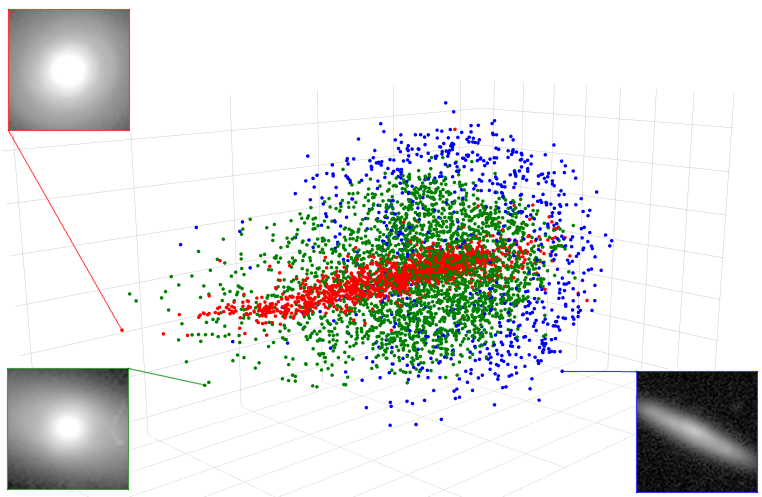} & \includegraphics[height=\myimageheight]{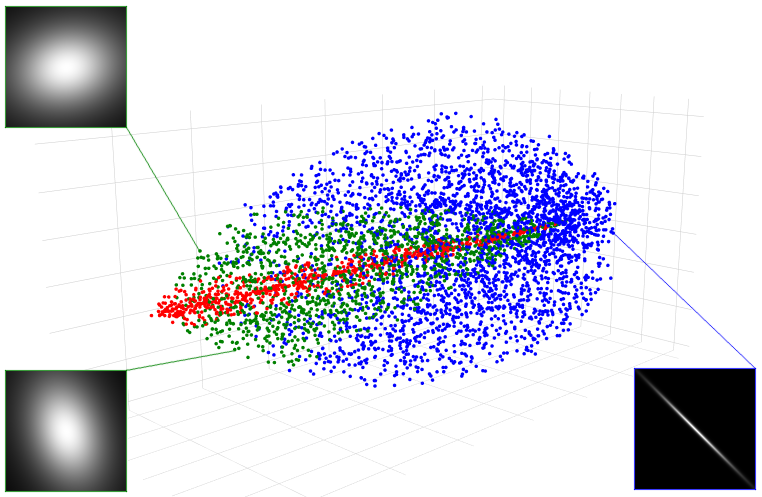} \\ 
GZD-5 Isomap (neighbors=30) & synthetic dataset Isomap (neighbors=30)\\
\hline
\includegraphics[height=\myimageheight]{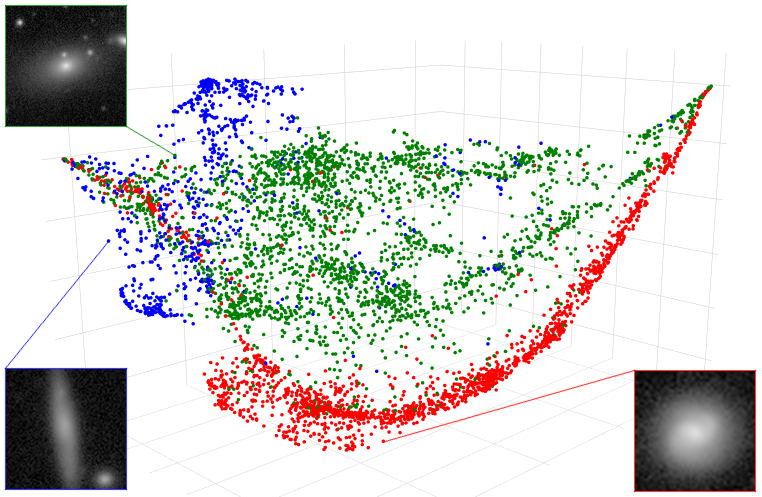} & \includegraphics[height=\myimageheight]{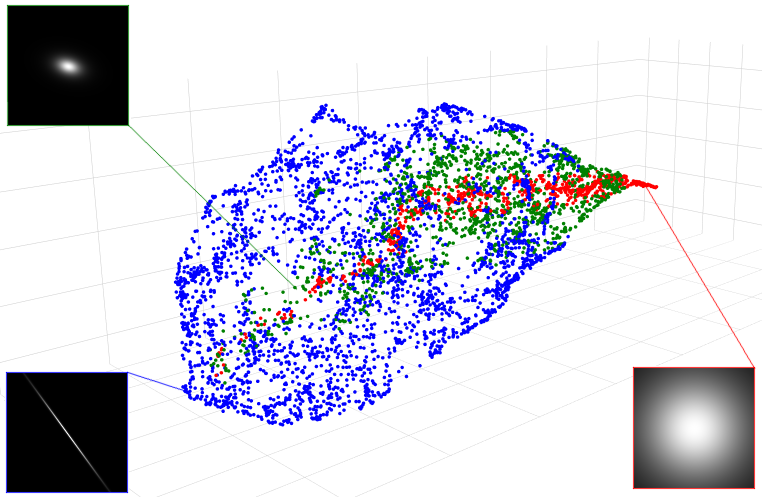} \\ 
GZD-5 UMAP (neighbors=30, min\_dist=0.1) & synthetic dataset UMAP (neighbors=30, min\_dist=0.1) \\
\hline
\includegraphics[height=\myimageheight]{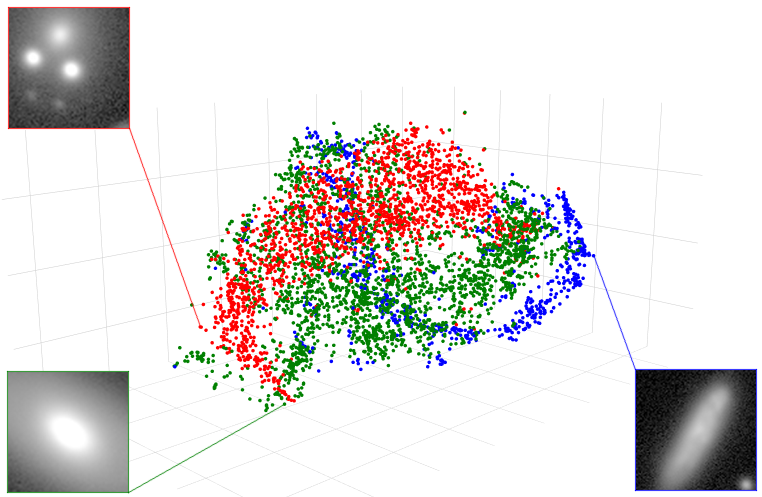} & \includegraphics[height=\myimageheight]{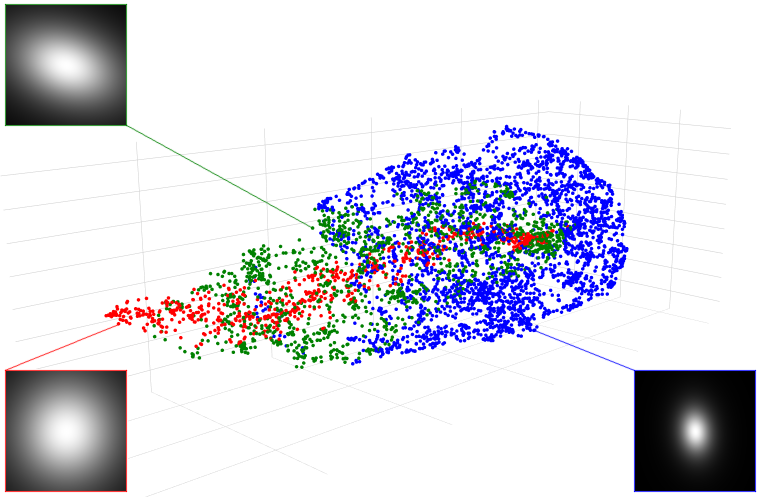} \\
GZD-5 t-SNE (perplexity=50) & synthetic dataset t-SNE (perplexity=50)\\
\hline
\includegraphics[height=\myimageheight]{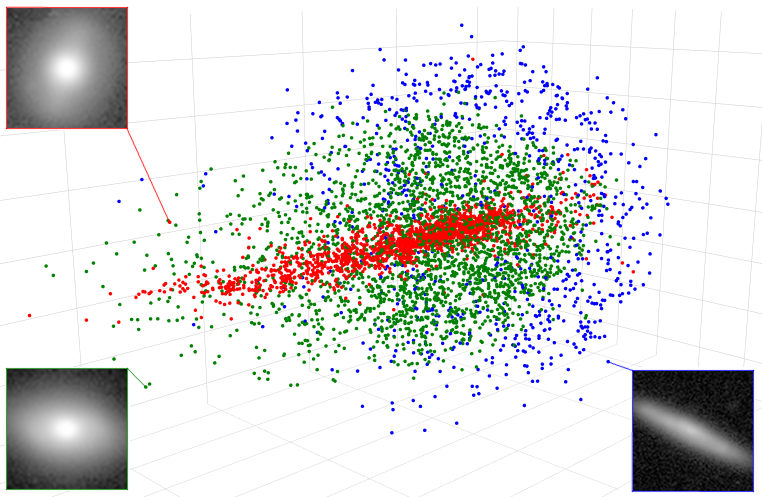} & \includegraphics[height=\myimageheight]{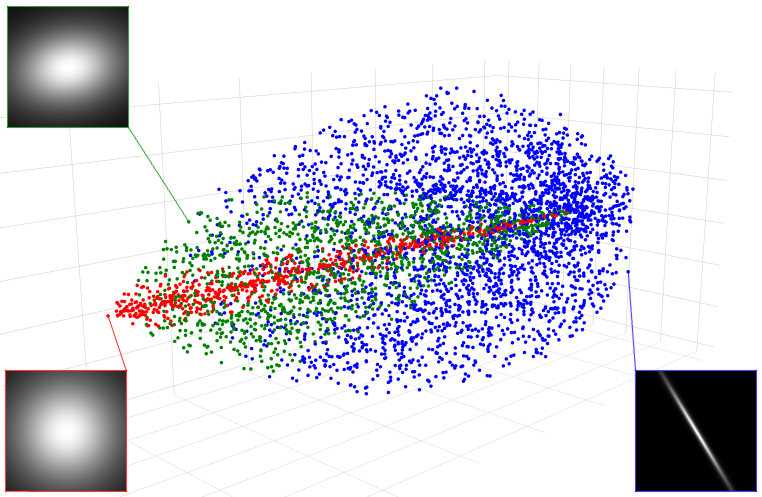} \\
GZD-5 LLE (neighbors=30) & synthetic dataset LLE (neighbors=30) \\
\hline
\end{supertabular}
\caption{Dimensionality reduction for GDZ-5 and synthetic data in 3D case. The method used and its parameters are defined under each figure.}
\label{tbl_3d_reduced}
\end{figure*}

It turns out that galaxy images are transformed into the points of an almond-shaped 3D manifold.
Moreover, the extracted dimensions have an interpretation, which is best understood in "conical" coordinates aligned with the axis of symmetry of the reconstructed structure.
In this framework, the azimuthal angle $\theta$ correlates with the third flattening measure (that could be expressed via eccentricity), the polar angle $\phi$ with the galaxy’s orientation angle, and the $z$-component with the total intensity or scale (product of the half-axis of the galaxy's elliptic profile) of the galaxy image.
The mapping also exhibits a spinoric property: a complete revolution of the polar angle results in a half-rotation of the galaxy image. The 3D manifold reconstructs an image of elliptical shapes in a manner similar to how the Poincaré sphere represents elliptically polarized light in optics.
The general scheme of our interpretation can be seen in figure~\ref{fig_manifold_scheme}, while our experimental results for all methods are presented in figure~\ref{tbl_3d_reduced}.

One can observe that LLE, Isomap, and PCA produce pretty similar results, while UMAP and t-SNE look a bit different even hinting at a separation of classes.
We have created a demo in JS that allows to rotate the manifold and see images of galaxies corresponding to different points~\cite{demo_galaxies}.

Additionally, we developed a three-parameter simple model to simulate the round/in-between/cigar dataset, providing insights into the three-dimensional manifold formed by real galaxies (see also \citep{euclid} as an example of training on synthetic galaxy morphology data). According to this model, a galaxy placed in the center of the $120\times 120$ image is characterized by its two axes and an orientation angle. The axis and the angles are sampled uniformly from the respective intervals. The galaxy’s profile is modeled using a generalized Schuster profile with an exponent of $-2.5$ although Gaussian or Sérsic profiles could be used as well. The labels of the generated galaxy images are assigned based on their eccentricity: round galaxies ($e \leq 0.5$), intermediate shapes ($0.5 < e \leq 0.8$), and elongated (cigar-shaped) galaxies ($e > 0.8$).
Experimental results indicate that setting the number of neighbors to 30 yields satisfactory outcomes for a dataset of $6,000$ images. A lower number of neighbors may lead to degenerate or unstable surfaces.

Another simple approach to generating elliptic shapes involves starting with a centered white circle on a black background, stretching it along the 
$x-$ and $y-$ axes of the image, and then rotating it. Alternatively, one could experiment with stretched squares as an approximation of ellipses. As previously mentioned, we have found an approximation that can be expressed in the Cartesian coordinate system as
$$
(a, b, \varphi) \mapsto \begin{pmatrix} k_1 \, a \, b \, \frac{a-b}{a+b} \,\cos{2 \varphi}\\
            k_1 \, a \, b \, \frac{a-b}{a+b} \, \sin{2 \varphi}\\
            k_2 \, a \, b
            \end{pmatrix},
$$
where $a$ is the larger semimajor axis, $b$ -- the smaller one, $\phi \in [0, \pi)$ - the angle between the $x$-axis of the image and the semimajor axis, $n=(a-b)/(a+b) = (1-\sqrt{1-e^2})/(1+\sqrt{1-e^2})$ is the so called third flattening measure (which has been used as a small parameter in Ramanujan's approximation and series expansions for the perimeter of an ellipse~\cite{linderholm1995overlooked}) and $k_1, k_2$ -- scaling parameters. 

LLE, Isomap, and PCA reconstruct an almond-shaped structure with points smoothly distributed along the figure (see fig.~\ref{tbl_3d_reduced}). This figure exhibits rotational symmetry, which corresponds to the rotational symmetry observed in the dataset images. The labels are determined by intersecting the reconstructed figure with cones oriented along its structure, with their vertices at the end corresponding to the smallest galaxies. This suggests a simple method for the classification of the model galaxies: use perspective projection onto the surface perpendicular to the main surface of the body and use SVC with RBF kernel. The model allows galaxies to be arbitrarily small. However, if a size cutoff is introduced, the resulting structure would be a truncated version of the almond-shaped manifold. In this case, orthogonal projection may be used instead of perspective projection to give decent results.

t-SNE and UMAP reconstruct a warped, shell-shaped version of the almond manifold. The structure obtained through LLE, Isomap, and PCA suggests that supervised classification methods may be effective, whereas the complex shapes produced by t-SNE and UMAP indicate potential suitability for unsupervised classification approaches as well.
We have also created a JS demo for this case~\cite{demo_model}.

\begin{figure}[htb]
\centering
\includegraphics[width=\linewidth]{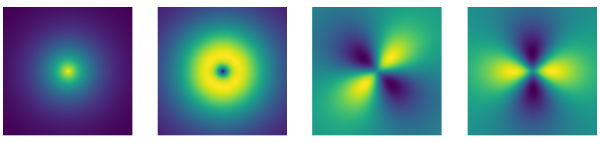}
\caption{Mean of the images (leftmost) and three first PCA components}
\label{fig:pca_components}
\end{figure}

The 3D structures reconstructed from the GZD-5 round/in-between/cigar dataset exhibit a high degree of similarity to those of the model galaxies. In the DECaLS dataset, the distribution of axis is not uniform: an average elliptical galaxy has an elongation of about 1.5. Therefore some areas of the 3D structure are undersampled. Since most galaxies in the dataset occupy a substantial portion of the image, dimensionality reduction algorithms are expected to reconstruct a truncated version of the almond manifold.
The interpretation of the components remains consistent across methods. Deviations mostly arise due to imperfections in the dataset; for example, a round galaxy located near a bright star is often misinterpreted by the models as a single elongated galaxy. Furthermore, images of small galaxies frequently contain numerous other luminous objects that confuse the models.

The principal components derived from PCA correspond to image features reminiscent of circular, diagonal, and linear polarization patterns or harmonics in figure~\ref{fig:pca_components}. The shape of the manifold created by PCA is once again the "almond" that we have seen before.

\section{Conclusion}
\label{conc}

In this study, we examine the problem of morphological classification of galaxies from the Galaxy Zoo DECaLS dataset using classical machine learning techniques. Our primary focus is on the dimensionality reduction step, which can be employed as a preprocessing stage in this framework.

First, we evaluated the performance of the most common dimensionality reduction methods, including Locally Linear Embedding (LLE), Isomap, Uniform Manifold Approximation and Projection (UMAP), t-SNE, and Principal Component Analysis (PCA), followed by a classical classification method (specifically SVC, logistic regression, and decision tree) to categorize galaxies based on their shape (cigar/in-between/ round; edge-on/face-on) and texture (smooth/featured).
For each combination of dimensionality reduction and classification methods, we conducted Bayesian optimization using the  \textit{optuna} library.
The results demonstrated that LLE consistently achieved the best performance, regardless of the classifier used, while the second-best method varied depending on the classification algorithm.

To better contextualize the obtained performance, we developed two simple neural networks for comparison: a convolutional neural network and a multi-layer perceptron. The results demonstrated that classical machine learning methods achieve performance comparable to these simple neural networks.

During the optimization process, we observed that performance exhibited only a weak dependence on the parameters of the dimensionality reduction methods. This suggests that only a small number of coordinates contain valuable information. Based on this finding, we reduced the dimensionality to three and generated the corresponding visual representations.
It was found that all methods produce the almond-shaped manifold, and, moreover, all components in the three-dimensional case exhibit a clear physical meaning. This suggests that all tested dimensionality reduction methods are highly interpretable. Notably, this high level of interpretability is unexpected, as nonlinear transformations typically result in a loss of interpretability.

To further investigate interpretability, we constructed a simple model to generate a synthetic dataset and applied all dimensionality reduction methods to this dataset. The results closely resembled those obtained from real galaxy images, reinforcing the consistency and robustness of the observed patterns.
In our model, the galaxy’s profile is modeled using a generalized Schuster profile with an exponent of $-2.5$ although Gaussian or Sérsic profiles could be used as well.
It was observed that the shape of the "almond" varies slightly depending on the chosen profile, suggesting potential avenues for further research to investigate the relationship between these parameters.

Lastly, we explored dimensionality reduction followed by k-means clustering to assess whether the data exhibits a natural tendency toward a specific number of clusters. 
We evaluate clustering performance using silhouette, elbow, Dunn, and Davies-Bouldin scores. 
While the Davies-Bouldin score indicates a slight preference for four clusters\,--- closely aligning with classifications made by human astronomers\,--- the other metrics do not indicate a clear clustering structure.

Hence, we can conclude that LLE and manifold learning in general can become a handy tool for problems of galaxy or, more generally, image morphological classification in astrophysics. There are some disadvantages of this method, in particular, it is not always obvious what dimensionality one should choose when reducing it, and often a hint from the outside method is needed. On the other hand, the advantages are more than clear: dealing with reduced data consumes much less computation power (each of the classification results can be achieved with only an hour of computation on a standard free Google Colab notebook, given the data was downloaded and pre-processed before), while the accuracy is competitive with other methods. The next steps in studying the application of manifold learning in this field could be more intense data filtration and preprocessing before reducing dimensionality and using more complex classification or clustering methods after.

The nearest future of galaxy observation is quite promising. Already functioning telescopes such as the Euclide \citep{euclid} and the James Webb Space Telescope \citep{jwst} are expected to, or are already giving an astonishing new level of detalization when observing morphologies of galaxies. Moreover, a new kind of morphology classification will be needed with start of the observation of SKA \citep{ska}. Therefore, developing diversified and miscellaneous classification pipelines for this problem is important. The one we propose here opens the way for mixing features from different wavelengths. Among other merits are the possibility of making fast unsupervised classifications of morphology without having a heavy pre-trained model. Hence, we see a deep potential for developing this approach in the future.

\section*{Acknowledgements} 
We would like to thank the Armed Forces of Ukraine for providing security to perform this work. This work was supported by the Ministry of Education and Science of Ukraine grant number 0122U001834 and by the National Academy of Sciences grant number 0120U101734.

\bibliographystyle{elsarticle-harv} 
\bibliography{bibliography}

\section*{Appendix}
Results of supervised classification for neural networks to have a benchmark for the results of our pipeline. We used the convolutional neural network (CNN, tables \ref{tab:cnn1report}, \ref{tab:cnn2report}, \ref{tab:cnn3report}) and fully-connected neural network (FCNN, tables \ref{tab:fcnn1report}, \ref{tab:fcnn2report}, \ref{tab:fcnn3report}). Results of {\textit {optuna}} search for optimal parameters for classification cases smooth/featured, and edge-on/face-on are in tables \ref{tbl_lle_params2}, \ref{tbl_lle_params3}.
\begin{table}[htb]
\centering
\begin{tabular}{lrrrr}
\toprule
 & precision & recall & f1-score & support \\
\midrule
round & 0.97 & 0.98 & 0.97 & 6,663 \\
in-between & 0.98 & 0.97 & 0.97 & 10,050 \\
cigar &  0.97 & 0.97 & 0.97 & 3,287 \\
\hline
accuracy & 0.97 & - & - & - \\
macro avg & 0.97 & 0.97 & 0.97 & 20,000 \\
weighted avg & 0.97 & 0.97 & 0.98 & 20,000 \\
\bottomrule
\end{tabular}
\caption{CNN results for round/in-between/cigar case.}
\label{tab:cnn1report}
\end{table}

\begin{table}[htb]
\centering
\begin{tabular}{lrrrr}
\toprule
 & precision & recall & f1-score & support \\
\midrule
smooth & 0.96 & 0.98 & 0.97 & 10,245 \\
featured & 0.97 & 0.96 & 0.95 & 9,755 \\
\hline
accuracy & 0.97 & - & - & - \\
macro avg & 0.97 & 0.97 & 0.97 & 20,000 \\
weighted avg & 0.97 & 0.97 & 0.97 & 20,000 \\
\bottomrule
\end{tabular}
\caption{CNN results for smooth/featured case.}
\label{tab:cnn2report}
\end{table}

\begin{table}[htb]
\centering
\begin{tabular}{lrrrr}
\toprule
 & precision & recall & f1-score & support \\
\midrule
edge-on & 0.94 & 0.95 & 0.95 & 3,489 \\
face-on & 0.99 & 0.99 & 0.99 & 16,511 \\
\hline
accuracy & 0.98 & - & - & - \\
macro avg & 0.97 & 0.97 & 0.97 & 20,000 \\
weighted avg & 0.98 & 0.98 & 0.98 & 20,000 \\
\bottomrule
\end{tabular}
\caption{CNN results for edge-on/face-on case.}
\label{tab:cnn3report}
\end{table}

\begin{table}[htb]
\centering
\begin{tabular}{lrrrr}
\toprule
 & precision & recall & f1-score & support \\
\midrule
round & 0.96 & 0.96 & 0.96 & 6,626 \\
inbetween & 0.95 & 0.96 & 0.95 & 10,045 \\
cigar &  0.97 & 0.90 & 0.93 & 3,329 \\
\hline
accuracy & 0.95 & - & - & - \\
macro avg & 0.96 & 0.94 & 0.95 & 20,000 \\
weighted avg & 0.95 & 0.95 & 0.95 & 20,000 \\
\bottomrule
\end{tabular}
\caption{FCNN results for round/in-between/cigar case.}
\label{tab:fcnn1report}
\end{table}
\begin{table}[!h]
\centering
\begin{tabular}{lrrrr}
\toprule
 & precision & recall & f1-score & support \\
\midrule
smooth & 0.79 & 0.92 & 0.85 & 10,237 \\
featured & 0.90 & 0.74 & 0.81 & 9,763 \\
\hline
accuracy & 0.83 & - & - & - \\
macro avg & 0.85 & 0.83 & 0.83 & 20,000 \\
weighted avg & 0.84 & 0.83 & 0.83 & 20,000 \\
\bottomrule
\end{tabular}
\caption{FCNN results for smooth/featured case.}
\label{tab:fcnn2report}
\end{table}
\begin{table}[!h]
\centering
\begin{tabular}{lrrrr}
\toprule
 & precision & recall & f1-score & support \\
\midrule
edge-on & 0.90 & 0.77 & 0.83 & 3,447 \\
face-on & 0.95 & 0.98 & 0.97 & 16,553 \\
\hline
accuracy & 0.95 & - & - & - \\
macro avg & 0.93 & 0.88 & 0.90 & 20,000 \\
weighted avg & 0.94 & 0.95 & 0.94 & 20,000 \\
\bottomrule
\end{tabular}
\caption{FCNN results for edge-on/face-on case.}
\label{tab:fcnn3report}
\end{table}

\begin{table}[!h]
\begin{tabular}{rlrrr}
\toprule
accuracy & method & components & neighbors & perplexity \\
\midrule
0.96 & LLE & 64 & 96 & N/A \\
0.92 & UMAP & 15 & 105 & N/A \\
0.94 & Isomap & 17 & 148 & N/A \\
0.83 & t-SNE & 2 & 93 & 30 \\
\bottomrule
\end{tabular}
\caption{Fine-tuned parameters for each of the dimensionality reduction methods used. The classification method in the pipeline is logistic regression, the classification case is smooth/featured}
\label{tbl_lle_params2}
\end{table}
\begin{table}[!h]
\begin{tabular}{rlrrr}
\toprule
accuracy & method & components & neighbors & perplexity \\
\midrule
0.96 & LLE & 199 & 199 & N/A \\
0.78 & UMAP & 57 &  48 & N/A \\
0.75 & Isomap & 102 & 37 & N/A \\
0.54 & t-SNE & 2 & 147 & 48 \\
\bottomrule
\end{tabular}
\caption{Fine-tuned parameters for each of the dimensionality reduction methods used. The classification method in the pipeline is logistic regression, the classification case is edge-on/face-on.}
\label{tbl_lle_params3}
\end{table}

\vfill

\end{document}